\documentclass[superscriptaddress, citeautoscript, floatfix, twocolumn, preprintnumbers]{revtex4-1}
\usepackage{graphicx}
\usepackage{textcomp}
\usepackage{subcaption}
\usepackage{siunitx}
\usepackage{bm}
\usepackage{epstopdf}

\begin{document}

\title{Morphology dependent optical anisotropies in the n-type polymer 
    P(NDI2OD-T2)}

\author{Steven J. Brown} 
\affiliation{Materials Department, University of
    California, Santa Barbara, California 93106, United States}

\author{Ruth A. Schlitz} 
\affiliation{SAGE Electrochromics, Inc., 2 Sage Way,
    Faribault, Minnesota 55021, United States}
\altaffiliation[Previous address:
  ]{Materials Department, University of California, Santa Barbara, California
    93106, United States}

\author{Michael L. Chabinyc}
\affiliation{Materials Department, University of
    California, Santa Barbara, California 93106, United States}

\author{Jon A. Schuller} 
\affiliation{Electrical and Computer Engineering
    Department, University of California, Santa Barbara, California 93106,
    United States} 
\email{jonschuller@ece.ucsb.edu}

\date{July 24, 2016}

\begin{abstract}

  Organic semiconductors tend to self-assemble into highly ordered and oriented
  morphologies with anisotropic optical properties. Studying these optical
  anisotropies provides insight into processing-dependent structural properties
  and informs the photonic design of organic photovoltaic and light-emitting
  devices. Here, we measure the anisotropic optical properties of spin-cast
  films of the n-type polymer P(NDI2OD-T2) using momentum-resolved absorption
  and emission spectroscopies. We quantify differences in the optical
  anisotropies of films deposited with distinct face-on and edge-on
  morphologies. In particular, we infer a substantially larger out-of-plane tilt
  angle of the optical transition dipole moment in high temperature annealed,
  edge-on films. Measurements of spectral differences between in-plane and
  out-of-plane dipoles, further indicate regions of disordered polymers in low
  temperature annealed face-on films that are otherwise obscured in traditional
  X-ray and optical characterization techniques. The methods and analysis
  developed in this work provide a way to identify and quantify subtle optical
  and structural anisotropies in organic semiconductors that are important for
  understanding and designing highly efficient thin film devices.

\end{abstract}

\maketitle

Organic semiconductors hold great promise in optoelectronic applications such as
organic photovoltaics (OPVs)\cite{mazzio_future_2014} and organic light emitting
diodes (OLEDs)\cite{hung_recent_2002} due to their ease of processing
(potentially leading to high-throughput and low-cost manufacture) and molecular
tunability. Organic semiconductors typically self-assemble into highly ordered
and oriented morphologies. As such, great strides have been made in
characterizing and optimizing morphologies\cite{delongchamp_molecular_2011,
  delongchamp_high_2007, treat_microstructure_2013, diao_solution_2013,
  kim_molecular_2013}, with a particular focus on the
electrical\cite{noriega_general_2013, venkateshvaran_approaching_2014} and
optical\cite{cabanillas-gonzalez_pump-probe_2011, spano_spectral_2010,
  schwartz_conjugated_2003} properties relevant to devices. As most morphologies
are highly oriented, it is important to study the variation of these properties
along different directions. For instance, anisotropic electrical properties of
organic semiconductors directly impact charge
transport\cite{sundar_elastomeric_2004, reese_high-resolution_2007,
  sirringhaus_two-dimensional_1999} and must be accounted for in device
design\cite{shah_effect_2009}. Optical spectroscopies such as
ellipsometry\cite{campoy-quiles_optical_2005, campoy-quiles_determination_2008},
polarized absorption\cite{tremel_charge_2014}, polarized
photoluminescence\cite{soci_anisotropic_2007}, and
Raman\cite{james_thin-film_2011, kotarba_anisotropy_2010}, similarly reveal
anisotropic optical properties related to the refractive index, absorption,
emission, and vibrational modes that significantly impact the design and
efficiency of light-emitting\cite{schmidt_comprehensive_2013,
  grell_polarized_1999} and photovoltaic\cite{jo_boosting_2015,
  grote_morphology-dependent_2013} devices.

Recently, momentum-resolved photoluminescence (mPL) measurements have provided
new insight into magnetic dipoles in atomic
systems\cite{taminiau_quantifying_2012}, intra- and inter-molecular excitons in
H-aggregates \cite{schuller_orientation_2013}, and waveguide exciton polariton
modes \cite{ellenbogen_exciton-polariton_2011}. Here, we extend these techniques
to study absorption as well as emission properties in highly ordered polymer
films. We study the molecule P(NDI2OD-T2) which adopts distinct 'edge-on' or
'face-on' orientations depending on processing conditions. By characterizing the
optical anisotropies of both morphologies, we determine the average orientation
of the transition dipoles, and resolve subtle differences in morphology (in both
crystalline and non-crystalline regions). These results reveal structural
features previously invisible to diffraction techniques and suggest ways to
increase device performance through film morphology optimization.

% Fig 1: Chemical structure + geometry and transition dipole of P(NDI2OD-T2)
\begin{figure}
  \begin{subfigure}[b]{0.45\textwidth}
          \includegraphics[scale=1]{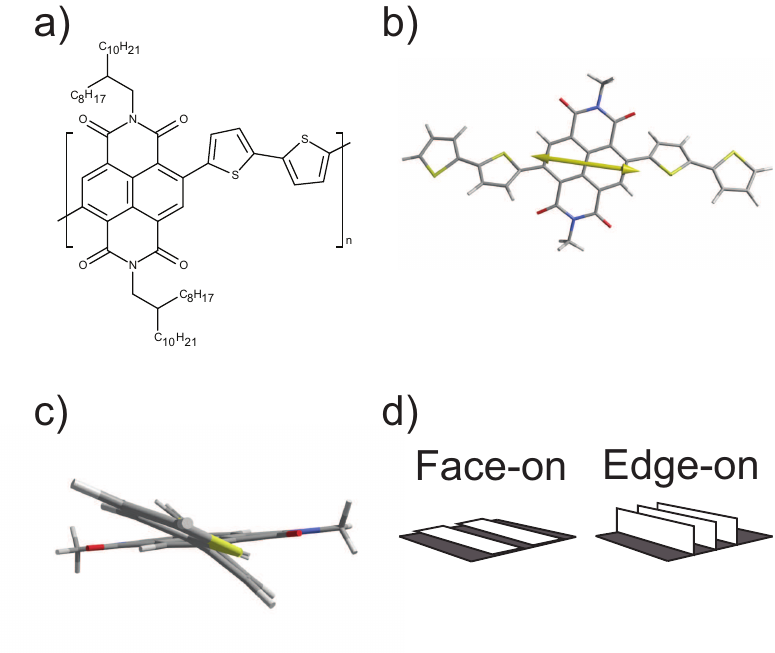}
  \end{subfigure}
  \caption{a)The chemical structure of P(NDI2OD-T2). b) Geometry and transition
    dipole moment of a P(NDI2OD-T2) molecule, determined with DFT calculations.
    The transition dipole moment (yellow arrow) lies in the plane of the NDI
    unit and is angled slightly with respect to the backbone. The alkyl
    side-chains have been truncated for visibility. c) A view down the backbone
    showing the relative twist of thiophene units with respect to the NDI2OD
    unit. While the subunits are twisted for both isolated and crystalline
    polymer, the exact angle in the solid state is dependent on morphology. d)
    When annealed at 150\textdegree C or 305\textdegree C P(NDI2OD-T2) takes on
    a face-on or edge-on morphology, respectively. The planes shown refer to the
    orientation of the NDI2OD planes.}
\end{figure}

P(NDI2OD-T2), sold by Polyera as N2200, is an n-type polymer. See Fig.\ 1(a) for
its structural formula and Fig.\ 1(b,c) for geometry. As n-type semiconducting
polymers are rare, P(NDI2OD-T2) has been the subject of extensive
morphology\cite{takacs_remarkable_2013, gann_near-edge_2014,
  giussani_molecular_2013, martino_mapping_2014, schuettfort_observation_2013,
  brinkmann_segregated_2012, anton_spatial_2016} and charge
transport\cite{fazzi_multi-length-scale_2015, caironi_very_2011,
  fabiano_orientation-dependent_2013, dinnocenzo_nature_2014,
  luzio_control_2013} studies.

P(NDI2OD-T2) is a particularly interesting system for studies of
structure-function relations because its molecular orientation can be controlled
through processing. Annealing films at a low temperature (150\textdegree C)
results in a face-on morphology where the pi-stacking direction is perpendicular
to the substrate (see GIWAXS, Supplemental Material, Fig.\ S1).\footnote[1]{See
  Supplemental Material at [URL will be inserted by publisher] for grazing
  incidence wide-angle X-ray scattering (GIWAXS) of face-on and edge-on
  P(NDI2OD-T2); ellipsometry of face-on and edge-on P(NDI2OD-T2).} In contrast,
a high temperature anneal (305\textdegree C) results in an edge-on morphology
with both the pi-stacking direction and polymer backbone parallel to the
substrate.\cite{rivnay_drastic_2011, schuettfort_surface_2011} These two
alternate morphologies are illustrated schematically in Fig.\ 1(d). Elucidating
the effects various morphologies have on film function and device performance is
on-going.

There have been a number of experiments linking processing conditions to optical
properties. Previous optical studies of P(NDI2OD-T2) in various solvents as well
as in thin films reveal subtle differences in absorption and photoluminescence
(PL) spectra depending on the degree of
aggregation.\cite{steyrleuthner_aggregation_2012, pavlopoulou_tuning_2014,
  kim_high-performance_2015} Subsequent studies used rubbing, directional
epitaxial crystallization, or epitaxy on oriented substrates to define a
preferential in-plane alignment of the polymer chains.\cite{tremel_charge_2014,
  brinkmann_segregated_2012} Polarized absorbance measurements then reveal
in-plane optical anisotropies: the films primarily absorb light with electric
fields polarized along the chain axis.\cite{tremel_charge_2014,
  brinkmann_segregated_2012} These studies demonstrate significant optical
structure-function relationships. However, these measurements of optical
anisotropies require specialized processing techniques to achieve in-plane
alignment and are insensitive to out-of-plane oriented optical properties. In
this paper we use momentum-resolved spectroscopies to measure the anisotropic
optical properties parallel vs. perpendicular to the substrate in films of
P(NDI2OD-T2) deposited with standard processing conditions and exhibiting no
preferred in-plane alignment over optical length scales.

\section{Results and Discussion}
 
  \subsection{Momentum-resolved spectroscopy}
  % Fig 2: Measurement geometry 
  \begin{figure}
    \begin{subfigure}[b]{0.45\textwidth}
      \includegraphics[scale=1]{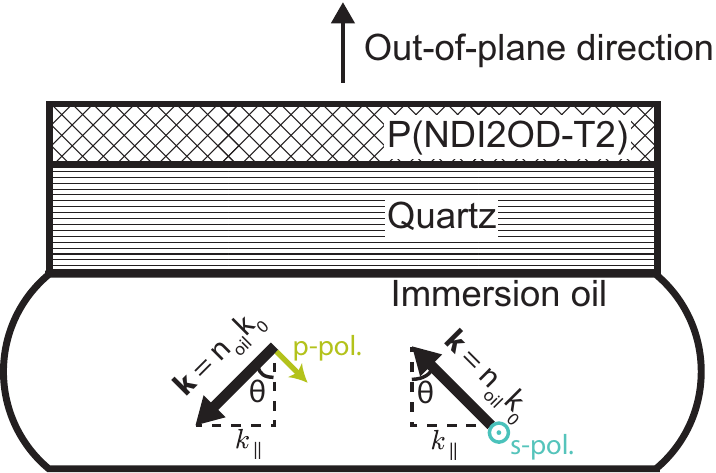}
    \end{subfigure}
    \caption{ Schematic showing the measurement geometry. In mPLE light is
      incident on the sample with a specific value of the in-plane momentum,
      $k_{||}$. In mPL light is emitted from the sample and measured as a
      function of $k_{||}$. The in-plane momentum is related to the angle of
      propagation as, $k_{||} = n_{oil}k_0\sin{\theta}$. Both p-polarized and
      s-polarized light are independently measured for both techniques.}
  \end{figure}
  % Fig 3: BFP images
  \begin{figure*}
  	\begin{subfigure}[b]{0.9\textwidth}
  		\includegraphics[scale=1]{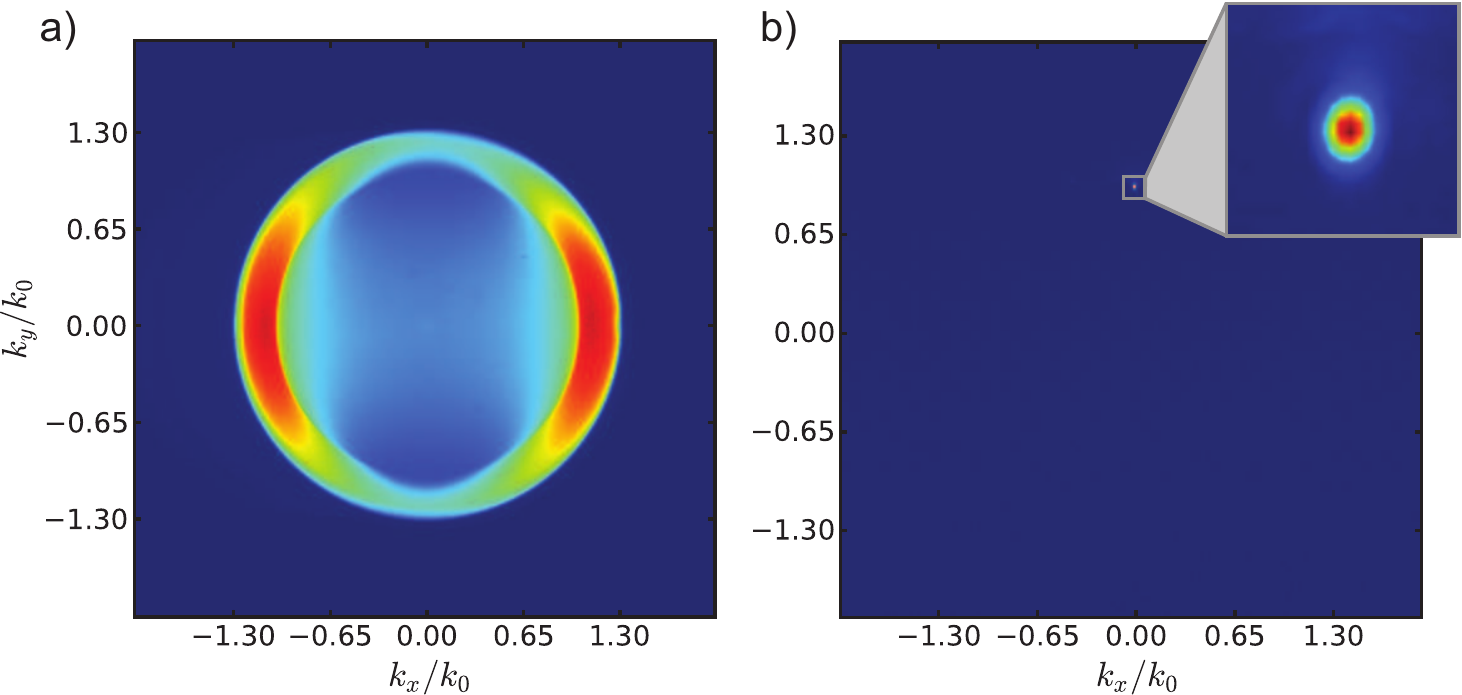}
  	\end{subfigure}
  	\caption{ (a) False color back focal plane image of y-polarized
      photoluminescence (750-1050nm integrated) from a P(NDI2OD-T2) film.
      Vertical (horizontal) linecuts through the center correspond to
      p-polarized (s-polarized) traces. (b) False color back focal plane image
      of reflected laser light, demonstrating momentum-resolved excitation at
      $k_x=-0.02 k_0$, $k_y=0.96 k_0$. By moving the output laser fiber within
      this plane we control the incidence momentum vector of our excitation
      source. The inset is a magnified image of the laser spot.}
  \end{figure*}

  % Fig 4: Pure IP/OP dipole linecut and decomposition
  \begin{figure*}
  	\begin{subfigure}[b]{0.9\textwidth}
  		\includegraphics[scale=1]{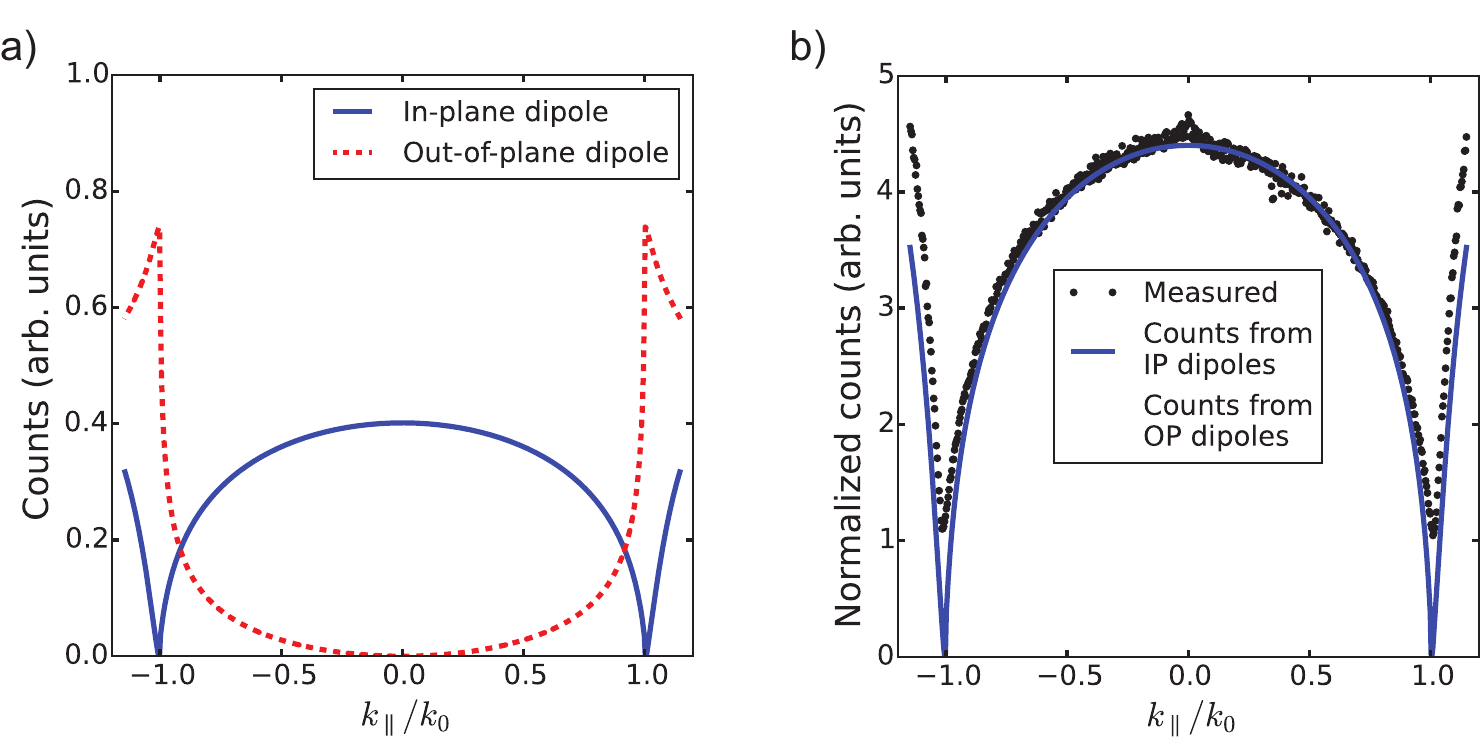}
  	\end{subfigure}
  	\caption{(a) Calculated momentum-dependent p-polarized luminescence expected
      from equal magnitude in-plane (blue, solid) and out-of-plane (red, dashed)
      emitting dipoles. (b) P-polarized photoluminescence at 865nm of
      P(NDI2OD-T2) annealed at 150\textdegree C decomposed into counts due to
      in-plane and out-of-plane dipoles.}
  \end{figure*}

  Momentum-resolved spectroscopies are a suite of techniques particularly
  well-suited to measuring the orientation of emitters, absorbers, and
  scatterers. In these techniques, variations in, e.g., PL, absorption,
  reflection, or scattered light intensity are measured as a function of the
  photon's momentum vector ($\bm{\vec{k}}$). These techniques utilize imaging in
  the back focal plane (Fourier plane) of a microscope objective (see Appendix
  A). Every point in the back focal plane corresponds to an angle
  of light incident on or emitted from the sample ($\theta =
  \arcsin{\frac{k_\parallel}{nk_0}}$, $\phi = \arctan{\frac{k_y}{k_x}}$ where
  $k_\parallel = \sqrt{{k_x}^2+{k_y}^2}$); see Fig.\ 2 for the measurement
  geometry. For example, in Fig.\ 3(a) we plot the p-polarized mPL from a thin
  film of P(NDI2OD-T2). Following previously established procedures,
  \cite{schuller_orientation_2013} we decompose mPL measurements like these into
  contributions from in-plane and out-of-plane oriented emission dipoles.
  Similarly, we also extend this basic technique to measure the orientation of
  \textit{absorption} dipoles. Incident light is focused to a point in the back
  focal plane, Fig.\ 3(b), such that it impinges the sample at a specific angle.
  By moving this focused light source within the back focal plane, we have
  complete control of the incident photon momentum vector and can measure
  momentum-resolved absorption. Both PL and absorption measurements are analyzed
  with simple electromagnetic models, detailed below.

  Because the emission (and absorption) distribution of a dipole is anisotropic
  ($\propto \sin^2\theta$), dipoles that are oriented in the plane of the sample
  (in-plane for the rest of the article) emit light into (or absorb light from)
  different angles than dipoles oriented perpendicular to the sample plane
  (out-of-plane). These differences are further amplified by reflections and
  interference in multi-layered geometries. Using a three-layer optical model,
  we calculate the p-polarized momentum-dependent PL intensity from purely
  in-plane (blue, solid) or out-of-plane (red, dashed) dipoles in a P(NDI2OD-T2)
  film, Fig.\ 4(a). The distributions are particularly different at normal
  incidence, where only in-plane (IP) dipoles emit, and at the critical angle,
  where only out-of-plane dipoles (OP) emit. We use these calculations to
  decompose measured momentum-resolved PL into contributions from IP and OP
  dipoles, Fig.\ 4(b). The 2D back focal plane PL image is focused to the
  entrance slit of an imaging spectrograph where it is separated spectrally. At
  each wavelength we measure the PL intensity as a function of in-plane momentum
  (black circles). The measured counts are decomposed into contributions from IP
  (blue, solid) and OP (red, dashed) dipoles. In this case, 5 percent of the
  total PL counts originate from OP dipoles.
 
  \subsection{Emission and absorption anisotropy}
  % Fig 5: Measured vs. calculated plots for mPL and mPLE
  \begin{figure*}
    \begin{subfigure}[b]{0.9\textwidth}
      \includegraphics[scale=1]{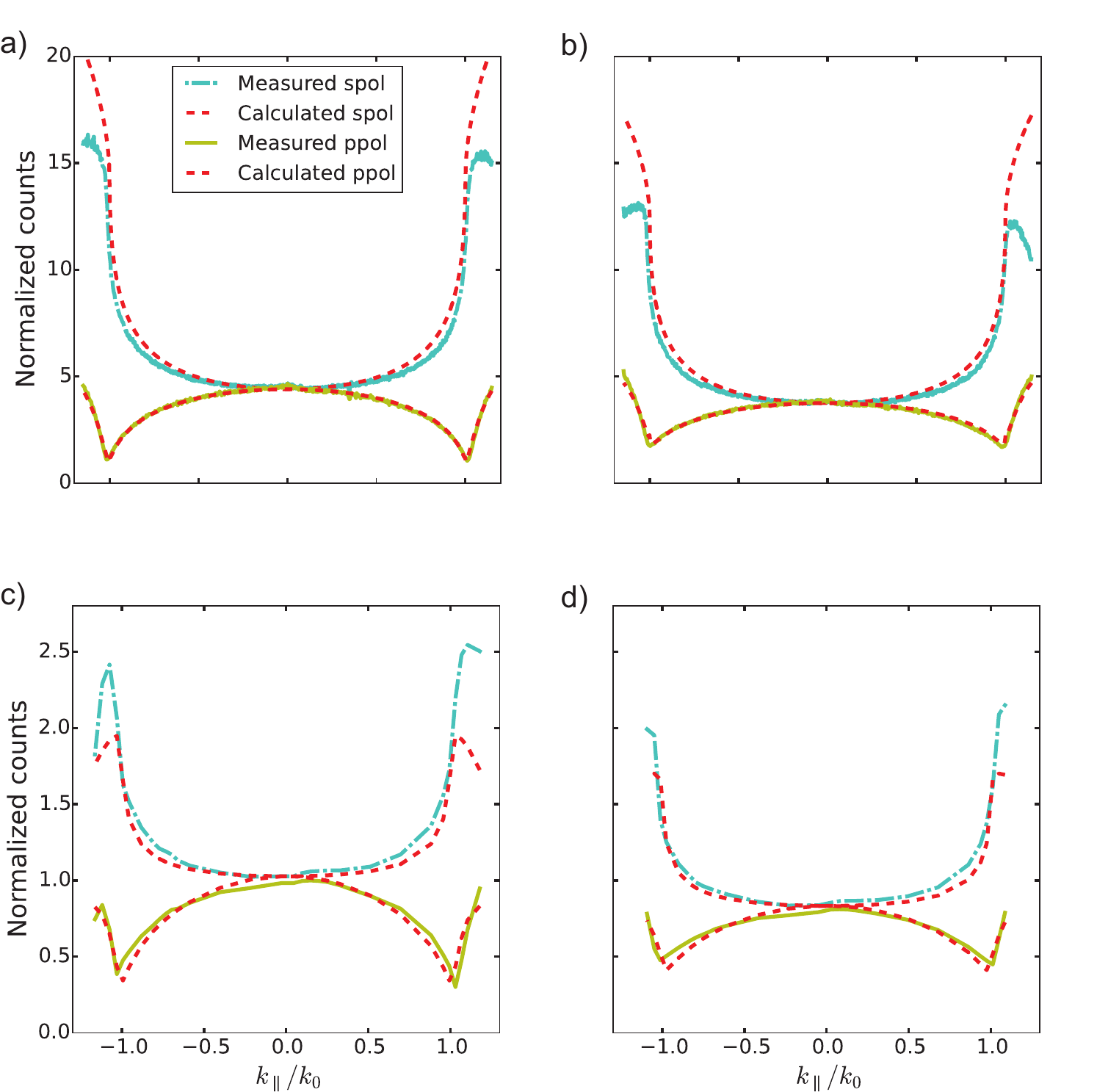}
    \end{subfigure}
    \caption{Examples of mPL measurements: 865nm photoluminescence intensity is
      recorded (solid, dot-dashed lines) vs.\ emission momentum for (a) face-on
      and (b) edge-on P(NDI2OD-T2). P-polarized traces (yellow, solid) are fit
      (dashed lines) to determine the relative contribution of IP and OP dipoles
      as illustrated in figure 4. From these fits we determine a predicted shape
      of the s-pol data (cyan, dot-dashed) with no free fit-parameters. The
      ratio of OP to IP dipole moments is 0.13 and 0.29 for face-on and edge-on
      films respectively. Examples of mPLE measurements: total photoluminescence
      intensity is recorded vs.\ incident momentum of 700nm excitation laser for
      (c) face-on and (d) edge-on P(NDI2OD-T2). The curves exhibit visible
      differences from mPL due to different values of the experimental
      apodization factor and values of the refractive index at 700nm vs. 865nm
      (see Appendix A). Regardless, the fit results of mPLE (0.16 and 0.30) show
      excellent agreement with mPL.}
  \end{figure*}

  Measured (solid, dot-dashed) and calculated (dashed) p-polarized (yellow,
  solid) and s-polarized (cyan, dot-dash) 865 nm PL traces for face-on and
  edge-on films are shown in Figs.\ 5(a,b). The s-polarized calculations contain
  no free fit parameters and show excellent agreement with measured PL up to
  approximately $k_\parallel = \pm 1.15*k_0$. This value of $|k_\parallel|$
  defines the range over which we perform fits of p-pol data\textemdash at
  larger momentum values the collection efficiency of the microscope objective
  begins to drop. The p-polarized experimental traces are fit according to the
  calculations described above, providing a measure of the relative contribution
  of in-plane and out-of-plane dipoles. As expected for excitations oriented
  primarily along the polymer chain, the emission in both morphologies is
  dominated by an in-plane dipole moment. However, fits of the p-polarized PL
  traces reveal a significant difference between the two morphologies. The ratio
  of out-of-plane to in-plane dipole moments is more than twice as large for
  edge-on (0.29) than face-on (0.13) morphologies. mPL allows us to resolve
  differences in the optical anisotropies that are \textit{not} evident in
  ellipsometry (Supplemental Material, Fig.\ S2).\footnote[1]{}

  From mPL we derive the orientation of the transition dipole moment (TDM) with
  respect to the substrate. The ratio of OP to IP dipole moments translate into
  differences of the average inclination angle of the TDM with respect to the
  substrate. The inferred angle is $\arctan(0.13) = \ang{7}$ for face-on films
  compared to $\arctan(0.29) = \ang{16}$ for edge-on orientations. This increase
  in angle is consistent with the orientation of NDI planes in face-on versus
  edge-on films. DFT calculations indicate a TDM is oriented mostly, but not
  completely, parallel to the polymer backbone. The TDM is tilted (8\textdegree)
  in the NDI plane, see Fig.\ 1(b).\cite{martino_mapping_2014,
    caironi_very_2011, fazzi_multi-length-scale_2015} In the edge-on
  morphology, the TDM is thus partially aligned perpendicular to the substrate.
  There may also be a tilting of the polymer backbone with respect to the
  substrate.\cite{giussani_structural_2015} and we cannot unambiguously identify
  the cause of differences in TDM orientation between the two morphologies. It
  is worth noting that the optical technique used here averages both crystalline
  and amorphous regions and therefore provides different information than can be
  found from X-ray diffraction alone.

  From Lorentz reciprocity\cite{potton_reciprocity_2004}, momentum-resolved
  light absorption is formally equivalent to momentum-resolved emission. Using
  the same principles described above, we provide the first demonstrations of
  momentum-resolved photoluminescence excitation (mPLE), a proxy for absorption
  (assuming photoluminescence intensity is linearly proportional to the amount
  of light absorbed). We collect the total emitted PL (integrated over
  wavelength and momentum) as a function of the position of our
  momentum-resolved laser excitation source, Fig.\ 2(b). Face-on and edge-on
  mPLE measurements are plotted in Figs.\ 5(c) and (d) at an excitation wavelength
  of 700 nm. Fitting these traces to the appropriate in-plane and out-of-plane
  basis functions at 700 nm (see Appendix C), we find out-of-plane to in-plane
  ratios of 0.16 and 0.30 for face-on and edge-on morphologies respectively.
  This excellent agreement with momentum-resolved emission (0.13 and 0.29)
  further validates our observation of larger TDM tilt-angles for edge-on
  polymer films. This also indicates minimal reorientation of the transition
  dipole between absorption and emission processes as can occur in other
  systems.\cite{montali_polarizing_1998, forster_redistribution_2007,
    schwartz_interchain_2001}
    
  \subsection{Spectral differences}
  % Fig 6: mPL compare dipole moments shape
  \begin{figure*}
    \begin{subfigure}[b]{0.9\textwidth}
      \includegraphics[scale=1]{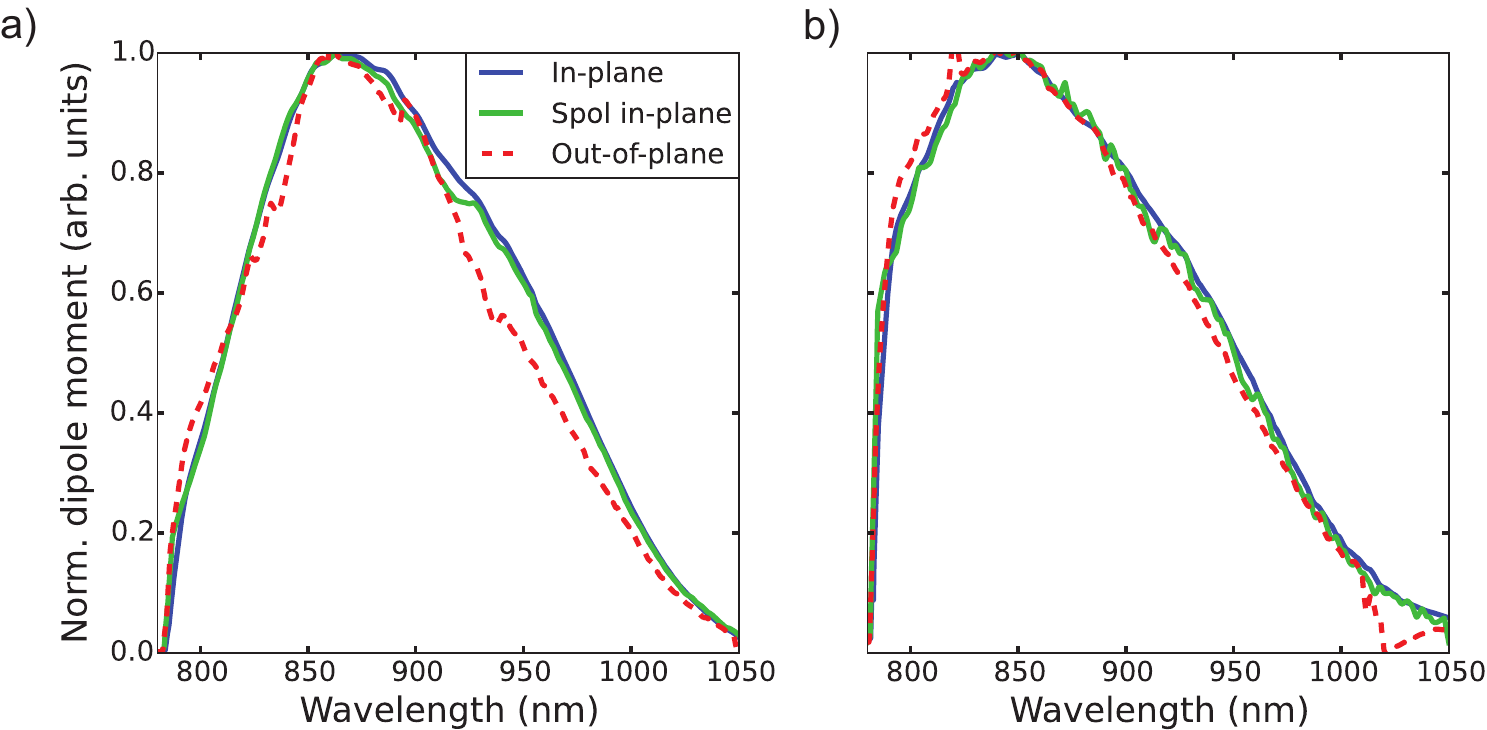}
    \end{subfigure}
    \caption{In-plane and out-of-plane normalized emission dipole moments for
      (a) face-on and (b) edge-on P(NDI2OD-T2), determined by performing mPL
      decompositions across the entire 780-1050 nm emission band.}
  \end{figure*}

  The wavelength dependence of these momentum-resolved measurements provides
  additional insight into the differences in optical properties for the two film
  morphologies. We only determine mPLE (i.e., absorption) at an excitation
  wavelength of 700 nm. The emitted light, on the other hand, is separated by
  momentum and wavelength simultaneously. Performing decompositions similar to
  Fig.\ 4(b), we observe an average OP/IP ratio of 0.12 with a standard deviation
  of 0.1 across the PL band (750-1050nm) for face-on films. Although the ratio
  is \textit{mostly} constant across the PL spectrum, deviations from these
  values are observed primarily at wavelengths to the right of the PL peak. This
  deviation is most easily visualized by plotting the normalized IP and OP
  spectra inferred from our fits at each wavelength. For face-on films, Fig.\
  6(a), the s-polarized spectrum (light green), which arises from only IP
  dipoles, is in close agreement with the IP spectrum determined from fits of
  p-polarized data (blue, solid). In particular, both spectra reveal a shoulder
  feature at 950 nm that is absent from the OP spectrum (red, dashed) determined
  from our fits. In comparison, edge-on films, Fig.\ 6(b) show much closer
  agreement between all three spectra (the out-of-plane artifact past 1000nm is
  due to low PL counts throwing off the fitting procedure). Evidently, the
  spectral dependence of these optical anisotropies reveals subtle differences
  in the morphology-dependent optical properties that are otherwise obscured.
  
  In previous studies, this 950nm shoulder peak was only seen in aggregated
  P(NDI2OD-T2).\cite{steyrleuthner_aggregation_2012}. A likely explanation for
  the missing face-on shoulder peak is that out-of-plane oriented dipoles are
  preferentially found in amorphous regions of the sample. When the polymer is
  initially spin-cast onto the substrate most of the molecules aggregate and
  align in the plane of the substrate. Some molecules, however, will exist in
  amorphous regions where there is a more random orientation of the molecules.
  The low temperature anneal provides only a small amount of energy for the
  molecules to rearrange and very little of the amorphous regions will
  crystallize. In this scenario out-of-plane dipoles are preferentially found in
  these randomly oriented amorphous regions. In contrast, a high temperature
  anneal, which gives rise to the edge-on morphology, has much more energy for
  the molecules to adjust and crystallize. A large portion of the amorphous
  polymers will crystallize while mostly retaining their original orientation.
  This simple model likely explains why the 950nm shoulder peak is found in the
  out-of-plane spectra for edge-on, but not face-on, morphologies.

\section{Conclusion}
In conclusion we use momentum-resolved spectroscopies to measure in-plane and
out-of-plane effective dipole moments for face-on and edge-on morphologies of
P(NDI2OD-T2). Fits of momentum-resolved emission measurements (mPL) show close
agreement with first-ever analogous absorption measurements using a
momentum-resolved photoluminescence excitation (mPLE) technique. We find that
edge-on films exhibit a larger out-of-plane tilt angle ($\sim$ 16\textdegree) of
the transition dipole moment relative to face-on films ($\sim$ 7\textdegree).
These results are consistent with the orientation of the transition dipole
moment within NDI planes, but may alternatively be indicative of a difference in
average orientation of the polymer backbones. We also observe a missing shoulder
peak, characteristic of aggregated P(NDI2OD-T2), in the out-of-plane emission
spectrum of face-on films. This suggests that the out-of-plane emission in these
films arises largely from amorphous regions. As typical optical techniques only
measure in-plane oriented dipoles and X-ray diffraction only measures
crystalline regions of the film, these out-of-plane amorphous regions have
likely been unexplored in previous studies. Finding annealing techniques that
maintain face-on orientation while crystallizing these previously hidden regions
will likely lead to better charge transport and, therefore, device performance
in organic photovoltaics and light-emitting diodes. In addition to these
insights on P(NDI2OD-T2)'s morphology, the momentum-resolved techniques
developed in this paper can be used to accurately characterize anisotropic
optical properties in other materials. These techniques can therefore enable new
optimizations of optical device design and reveal subtle differences in
morphology that are obscured in other X-ray and optical characterization
techniques.

\begin{acknowledgements}
  This work was supported by a National Science Foundation CAREER award
  (DMR-1454260). AFM and UV-Vis spectroscopy were performed in the MRL Shared
  Experimental Facilities which are supported by the MRSEC Program of the NSF
  under Award No.\ DMR 1121053; a member of the NSF-funded Materials Research
  Facilities Network (www.mrfn.org). A portion of this work was performed in the
  UCSB Nanofabrication Facility.
\end{acknowledgements}

 \appendix 
  \section{Experimental setup}
  % Fig 7: Experimental setup 
  \begin{figure*}
    \begin{subfigure}[b]{0.9\textwidth}
      \includegraphics[scale=1]{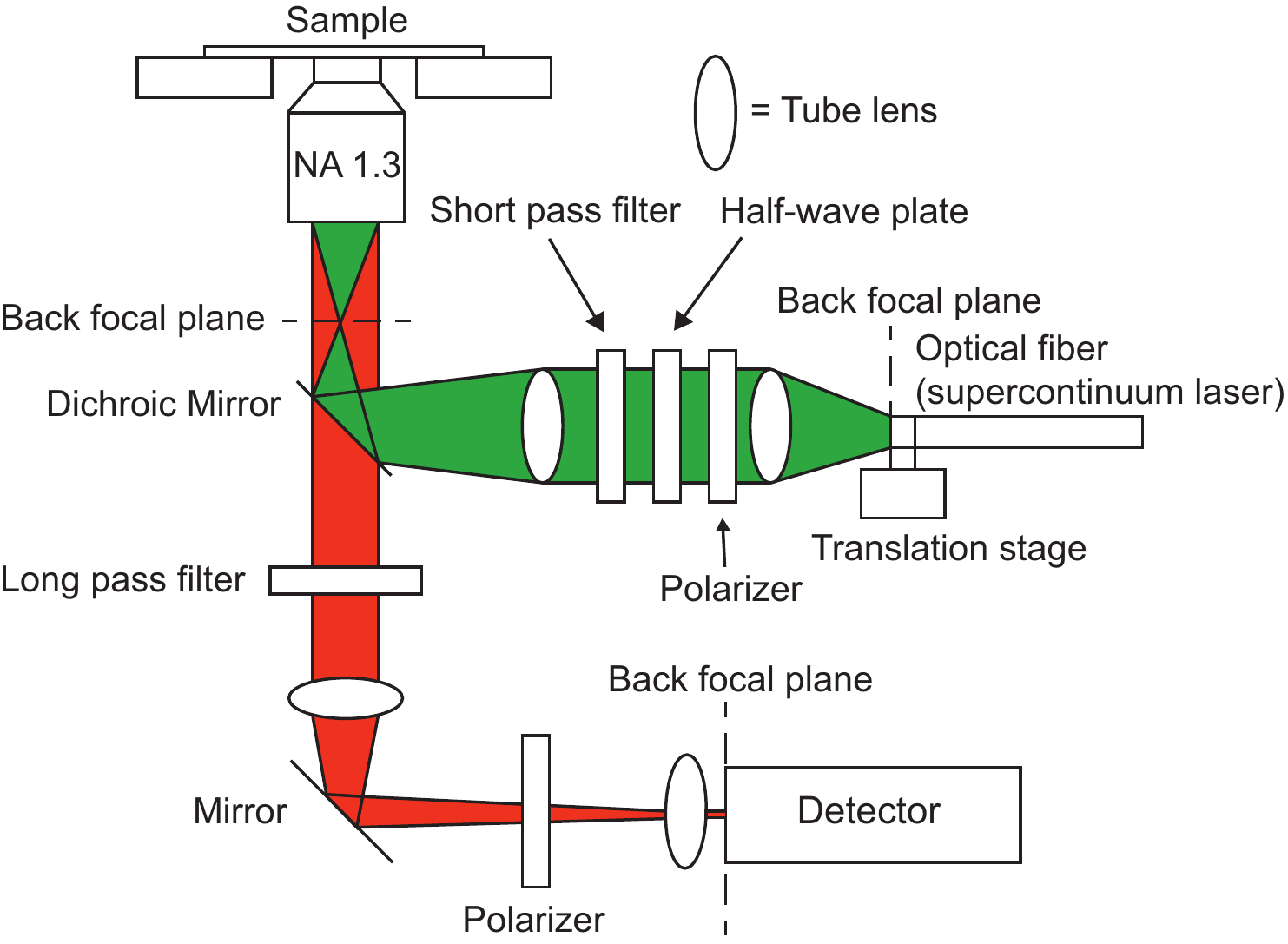}
    \end{subfigure}
    \caption{The momentum-resolved setup used in the experiments. For
      momentum-resolved photoluminescence, the supercontinuum
      excitation source along with the half-wave plate, polarizer, and first
      tube lens were replaced with a collimated LED.}
  \end{figure*}
  By placing a detector in the back focal plane of a microscope objective, Fig.\
  7,\footnote[1]{} we separate light based on the angle, or momentum, at which
  it leaves the sample. A spectrometer (Princeton Instruments IsoPlane SCT320)
  coupled to a 2D CCD camera (Princeton Instruments PIXIS 1024BRX) separates
  light by wavelength along one axis of the camera and momentum along the other
  axis. This allows measurements of momentum-dependent photoluminescence
  intensity at many wavelengths simultaneously. From this data, we separate
  emission spectra from dipoles oriented in-plane and out-of-plane. For mPL
  experiments we used a collimated LED source (ThorLabs M735L3-C5) to excite the
  sample across all momenta uniformly.

  Similarly to how placing a detector in a conjugate back focal plane to the
  objective allowed us to study emission of different momenta of light, we
  placed our excitation source in another conjugate back focal plane to study
  absorption as a function of light momenta. We studied absorption properties by
  measuring the integrated intensity of photoluminescence emitted from the
  sample versus the input excitation momentum (similar to how photoluminescence
  excitation, or PLE, measures PL versus input wavelength of light). The end of
  a single mode optical fiber (coming from a fiber-coupled supercontinuum source
  (SuperK Extreme EXR-15)) was mounted on a translation stage. By moving the end
  of the fiber within the conjugate back focal plane, we control the incident
  momentum vector of the light exciting the sample.

  \section{Sample fabrication}
  P(NDI2OD-T2) (Polyera ActivInk N2200) was spin-cast from 1,2-dichlorobenzene
  solution (\SI{10}{\micro\gram/\milli\liter}). Samples were then annealed at
  \SI{150}{\degreeCelsius} for one hour to produce face-on samples or at
  \SI{305}{\degreeCelsius} for one hour to produce edge-on samples. The samples
  were then allowed to slowly cool to room temperature. Film thickness was
  measured using atomic force microscopy (AFM). Glass, 200nm silicon dioxide on
  silicon, and quartz coverslip substrates were used for AFM, ellipsometry, and
  PL measurements respectively.

  \section{Data analysis}

  Raw camera images were analyzed using Python. In mPL we obtain a rectangular
  image with wavelength varying along the x-axis and momentum varying along the
  y-axis. To obtain in-plane and out-of-plane emission dipole moments we
  analyzed each wavelength independently. For a given wavelength column in the
  p-polarized image, we first converted the pixels of the camera into units of
  $k_0$ by setting the edges of PL to the 1.3 NA ($\pm 1.3k_0$) of our
  microscope objective. After converting to units of momentum we used a three
  layer model (following the treatment in Schuller \emph{et
    al.}\cite{schuller_orientation_2013}) to solve for the linear combination of
  in-plane and out-of-plane effective emission dipole moments that summed to the
  intensity vs. momentum shape measured at each wavelength. The model's input
  parameters---refractive index and film thickness---were determined from
  ellipsometry and AFM measurements respectively. After fitting p-polarized
  data, we determine the expected s-polarized PL intensity vs. momentum and
  compare to actual s-polarized measurements.
  
  Data analysis for mPLE was similar, but had many separate images that needed
  to be aggregated. For each PL image the exciting laser y-position was
  determined by taking the image of the reflected laser spot without the PL
  filter. Each PL image was background subtracted using a "window frame" of dark
  pixels that surrounded the pixels receiving PL in the center of the image.
  This allowed us to correct for background drift over time. The background
  subtracted PL image was then summed across all pixels to determine a single PL
  value for each image. From the PL image we were able to convert pixels to
  $k_0$ as above. We then found the linear combination of in-plane and
  out-of-plane effective absorption dipole moments that summed to the counts vs.
  momentum shape observed. Again, we determine the expected s-polarized PL
  intensity vs. incident momentum and compare to actual s-polarized
  measurements. It is important to note that while the mPL model includes an
  apodization factor, given the setup geometry the mPLE model does not. In mPL,
  each pixel of fixed width in the back-focal-plane correspond to a different
  magnitude of solid-angle over which the PL is collected. Thus, an isotropic
  emitter would still exhibit intensity variations across the back focal plane
  image. In mPLE, the solid-angle magnitude also changes, but the input power is
  fixed and no correction is needed.

  Dipole moments found via mPL and mPLE are highly sensitive to the film
  refractive indices input in the three-layer model (especially in the
  out-of-plane direction). For this reason it is essential to have accurate
  optical constants. We used atomic force microscopy, UV-Vis transmission, and
  ellipsometry to get accurate thickness, in-plane extinction coefficient, and
  refractive indices respectively.
  
  %

%%%%%%%%%% Supplemental materials %%%%%%%%%%
\clearpage
\widetext
\begin{center}
    \textbf{\large Supplemental Materials: Morphology dependent optical
         anisotropies in the n-type polymer P(NDI2OD-T2)}
\end{center}
%%%%%%%%%% Prefix a "S" to all equations, figures, tables and reset the counter %%%%%%%%%%
\setcounter{equation}{0}
\setcounter{figure}{0}
\setcounter{table}{0}
\setcounter{page}{1}
\makeatletter
\renewcommand{\theequation}{S\arabic{equation}}
\renewcommand{\thefigure}{S\arabic{figure}}
\renewcommand{\bibnumfmt}[1]{[S#1]}
\renewcommand{\citenumfont}[1]{S#1}
%%%%%%%%%% Prefix a "S" to all equations, figures, tables and reset the counter %%%%%%%%%%
\maketitle

\begin{figure}[h!]
    \begin{subfigure}[b]{0.9\textwidth}
        \includegraphics[scale=0.3]{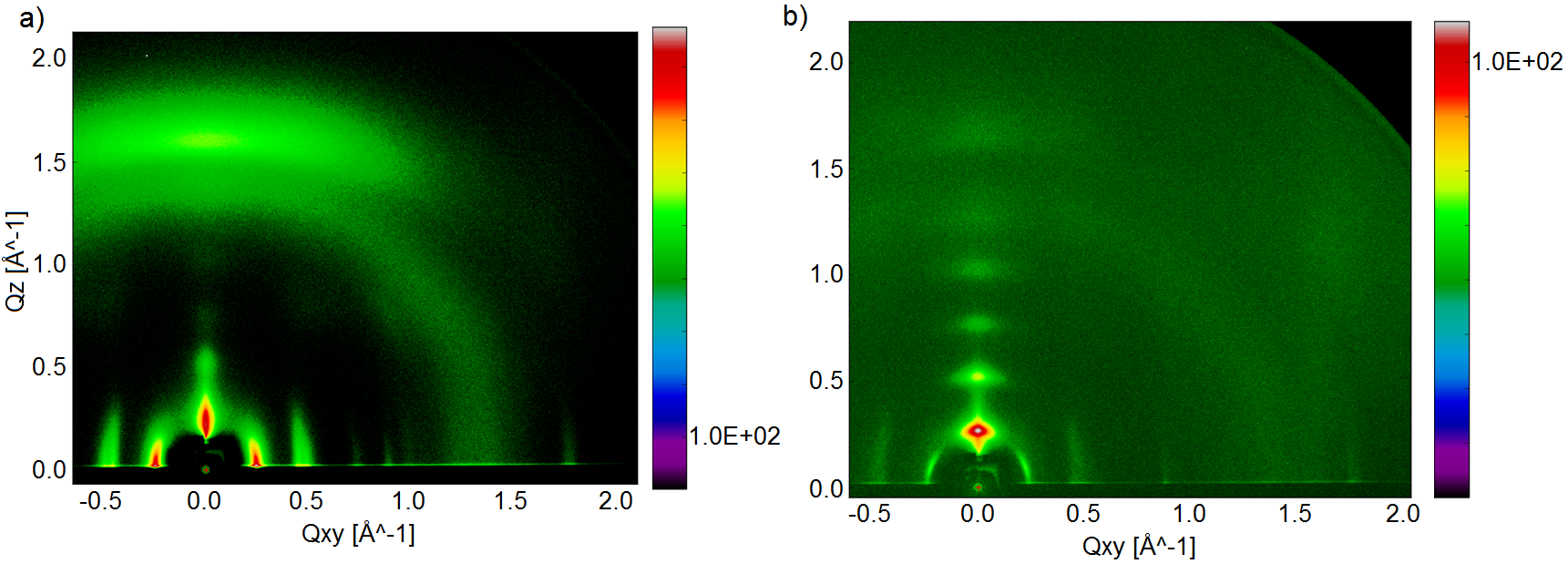}
    \end{subfigure}
    \caption{a)Grazing incidence wide-angle X-ray scattering (GIWAXS) for a
        face-on oriented P(NDI2OD-T2) annealed at 150\textdegree C for 6 hours and
        b)for an edge-on oriented P(NDI2OD-T2) annealed at 300\textdegree C for 45
        minutes.}
\end{figure}

\begin{figure}[h!]
    \begin{subfigure}[b]{0.9\textwidth}
        \includegraphics[scale=1]{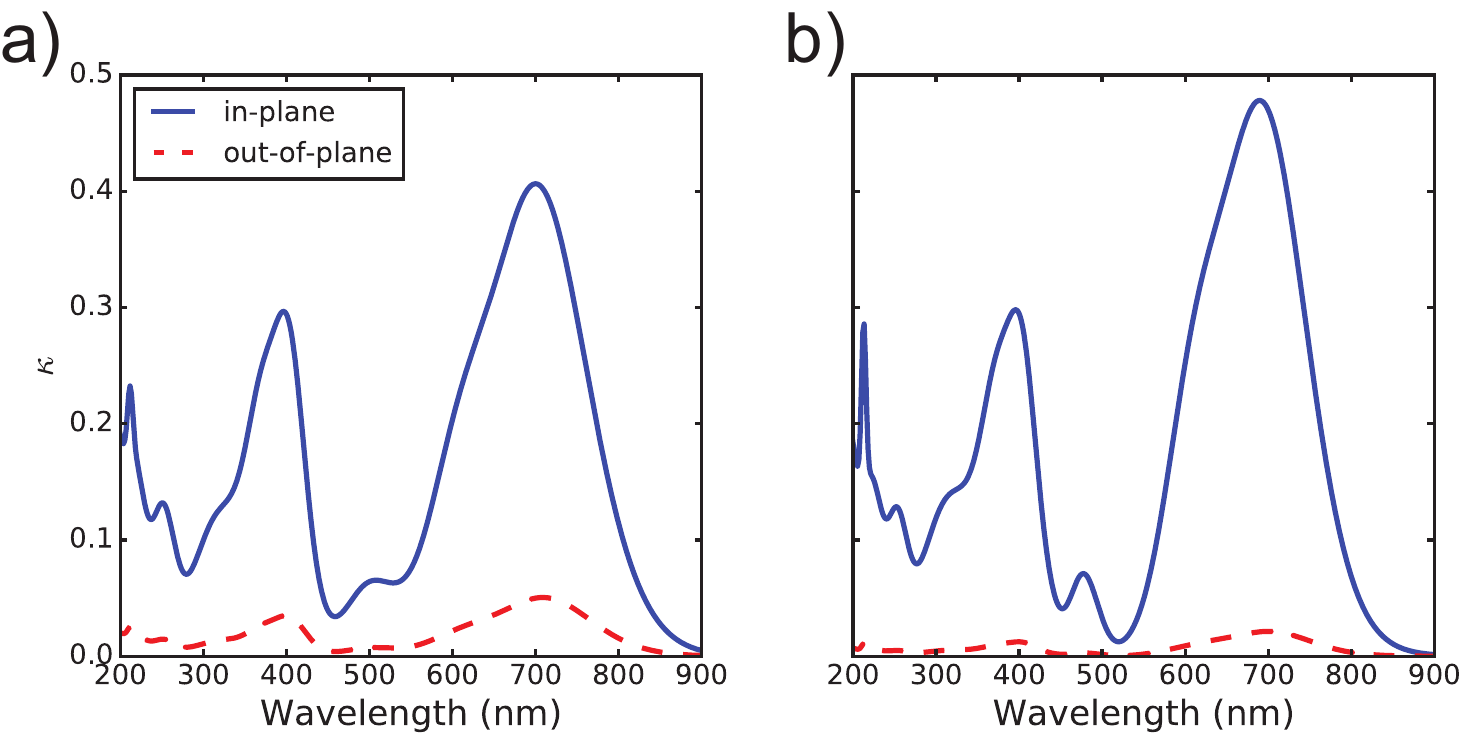}
    \end{subfigure}
    \caption{The extinction coefficient in the in-plane and out-of-plane
        directions (with respect to the substrate) for (a) face-on and (b) edge-on
        P(NDI2OD-T2), determined from spectroscopic ellipsometry measurements. The ellipsometry 
        results are highly model dependent and unable to capture the differences in optical anisotropies that 
        are evident in momentum-resolved measurements.}
\end{figure}

\end{document}